\documentclass[twocolumn,prb,showpacs,citeautoscript,floatfix]{revtex4}
\usepackage{array}
\usepackage{multirow}
\usepackage{amssymb}
\usepackage{graphicx}
\input{epsf}

\begin{document}

\title{Stabilization of carbon nanotubes by filling with inner tubes: An optical spectroscopy study on double-walled carbon nanotubes under hydrostatic pressure}
\author{B.~Anis$^{1}$,~K. Haubner$^{2}$,~F. B\"orrnert$^{2}$,~L. Dunsch$^{2}$,~M. H. R\"ummeli$^{2}$, and~C.~A.~Kuntscher$^{1}$}
\email[E-mail:~]{christine.kuntscher@physik.uni-augsburg.de}
\affiliation{$^{1}$~Experimentalphysik~2,~Universit\"at~Augsburg,~
D-86195~Augsburg,~Germany }
\affiliation{$^{2}$~IFW Dresden, P.O. Box 270116, D-01171 Dresden, Germany}

\date{\today}

\begin{abstract}
The stabilization of carbon nanotubes via the filling with inner tubes is demonstrated by probing the optical transitions in double-walled carbon nanotube bundles under hydrostatic pressure with optical spectroscopy. Double-walled carbon nanotube films were prepared from fullerene peapods and characterized by HRTEM and optical spectroscopy. In comparison to single-walled carbon nano\-tubes, the pressure-induced redshifts of the optical transitions in the outer tubes are significantly smaller below $\sim$10~GPa, demonstrating the enhanced mechanical stability due to the inner tube already at low pressures. Anomalies at the critical pressure P$_d$$\approx$12~GPa signal the onset of the pressure-induced deformation of the tubular cross-sections. The value of P$_d$ is in very good agreement with theoretical predictions of the pressure-induced structural transitions in double-walled carbon nanotube bundles with similar average diameters.
\end{abstract}


\maketitle

\section{Introduction}
Carbon nanotubes possess superior mechanical properties among carbon fiber materials due to the robust carbon atom network based on strong carbon-carbon covalent bonds.\cite{Jorio08} Their mechanical stability has been extensively investigated by probing their vibrational and structural properties under high external pressure via Raman spectroscopy and x-ray diffraction, respectively. These investigations demonstrated that for single-walled carbon nanotubes (SWCNTs) applied pressures of a few GPa lead to severe deformations of the nanotubes' circular shape and their collapse at high pressure.\cite{Lebedkin06,Tang00,Sharma01,Yao08,Caillier08}
The possible stabilization of SWCNTs against deformation by filling the empty tubes with molecules or with inner tubes is an important issue. High-pressure Raman measurements showed that in the case of double-walled carbon nanotubes (DWCNTs) the inner tube supports the outer one up to high pressure before the collapse.\cite{Aguiar11,Arvanitidis05} On the contrary, iodine and C$_{70}$ fillers are supposed to cause a destabilization of the SWCNTs.\cite{Caillier08,Aguiar11}

Up to now, the stability of SWCNTs and DWCNTs against hydrostatic pressure has been mainly addressed by Raman spectroscopy, which monitors the vibrational properties. Surprisingly, pressure experiments on carbon nanotubes using techniques probing the electronic structure, like optical spectroscopy, are rare and mostly cover only the low-pressure regime.\cite{Kazaoui00,Minami01,Wu04,Thirunavukkuarasu10}
In general, optical spectroscopy is a powerful technique to characterize the electronic band structure in terms of the energy position and spectral weight of the excited interband and intraband transitions. As demonstrated recently, the optical response is capable of monitoring
small pressure-induced deformations of the tubular cross-section, since the electronic band structure of SWCNTs is very sensitive to such deformations.\cite{Thirunavukkuarasu10,Kuntscher10} Here, we present the results of the first optical spectroscopy study on DWCNTs subjected to high hydrostatic pressure. The goal of this study was to test the mechanical stability of a carbon nanotube when filled with an inner tube in comparison with an empty tube, via the pressure-induced changes in the optical response.

\section{Experiment}

\textit{Chemicals}. Bundled SWCNTs were purchased from Carbon Solutions Inc.(Type P2, average
diameter 1.4~nm and batch No.~02-444). The P2 SWCNTs were prepared
using the arc discharge method. C$_{60}$-fullerene with purity 99.98\% was
purchased from Term USA. Triton X-100 ($\sim{10\%}$~in~H$_{2}$O) was purchased from Sigma-Aldrich.

\textit{Synthesis}. DWCNTs were prepared from SWCNTs $@$ C$_{60}$ peapods.
Arc discharge SWCNTs (1.4~nm average diameter, Carbon Solutions Inc.) were filled with C$_{60}$ molecules using the sublimation method \cite{B.W.Smith2002}. The prepared peapods were then transformed to DWCNTs by tempering the peapods at 1250$^{\circ}$C for 24~hrs under dynamic vacuum and subsequently cooling the furnace to room temperature. Free-standing films from SWCNTs and DWCNTs for optical spectroscopy measurements were prepared from Triton X-100 suspension. \cite{Z.Wu04}

\textit{TEM characterization.}
The samples were characterized in different preparation stages with a JEOL JEM-2010F transmission electron microscope retrofitted with two CEOS third-order spherical aberration correctors for the objective lens (CETCOR) and the condenser system (CESCOR). The microscope was operated using an electron acceleration voltage of 80 kV to reduce knock on damage.

\textit{Optical spectroscopy at ambient and high pressure}. Transmittance spectra of the nanotube films were measured at room temperature in the energy range 3500-22000 cm$^{-1}$ with the resolution 4~cm$^{-1}$ using a Bruker IFS 66v/S Fourier transform spectrometer in combination with an infrared microscope (Bruker IR Scope II) with a 15$\times$ Cassegrain objective. Syassen-Holzapfel type~\cite{Syassen1977} diamond anvil cell (DAC) was used for generation of pressure. Ruby luminescence technique was used for pressure measurement. Liquid nitrogen served as hydrostatic pressure transmitting medium. The intensity I$_{s}$($\omega$) of the radiation transmitted through the sample and the intensity I$_{ref}$($\omega$) of the radiation transmitted through the pressure transmitting medium in the DAC were measured. From I$_{s}$($\omega$) and I$_{ref}$ ($\omega$) the transmittance and absorbance spectra were calculated according to T($\omega$)=I$_{s}$($\omega$)/I$_{ref}$ ($\omega$) and A($\omega$)= -log$_{10}$T($\omega$), respectively.

\section{Results and discussion}

\subsection{Electron Microscopy}

\begin{figure}
\includegraphics[width=1.0\columnwidth]{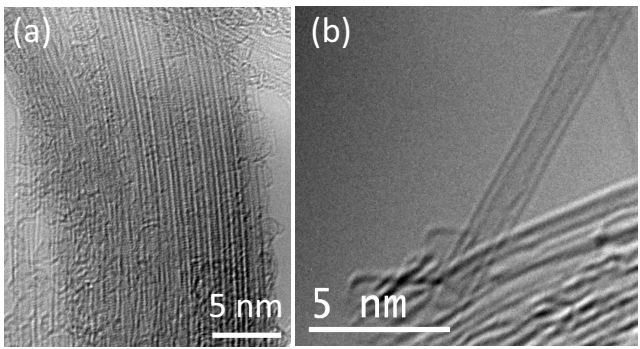}
\caption{(a) HRTEM image of a bundle of DWCNTs derived from C$_{60}$ peapods. (b) HRTEM image of a single DWCNT.} \label{fgr:HRTEM}
\end{figure}

Figure \ref{fgr:HRTEM} (a) shows one HRTEM image for the DWCNTs prepared
from the C$_{60}$ peapods after heat treatment at 1200$^{\circ}$C for 24~hrs.
To confirm the formation of inner tubes a single DWCNT is depicted in Fig.~\ref{fgr:HRTEM}(b). Based on the HRTEM images we estimate a filling ratio of $>$95\%.

\subsection{Optical spectroscopy on free-standing SWCNTs and DWCNTs films}

\begin{figure}
\includegraphics[width=0.9\columnwidth]{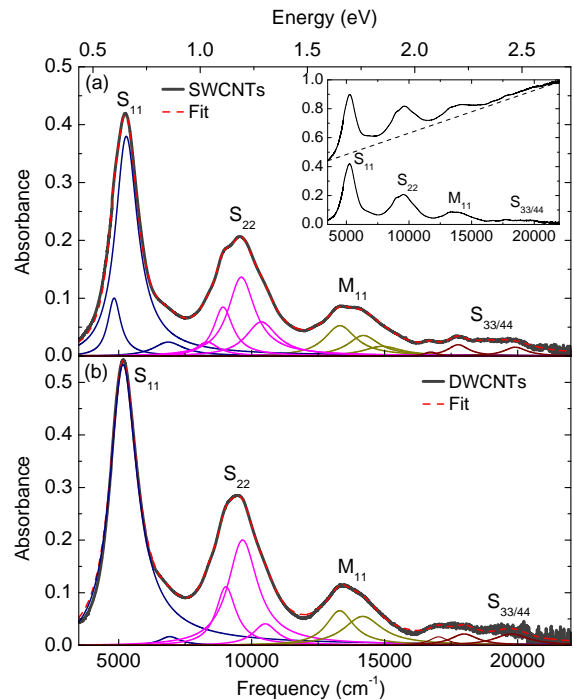}
\caption{Background-subtracted optical absorbance spectra of the free-standing film of (a) SWCNTs and (b) DWCNTs together with the fit of the absorption bands using Lorentzian oscillators.
The labels S$_{ii}$ and M$_{ii}$ denote the $i$th optical transition in semiconducting and metallic tubes,
respectively. Inset: Absorbance spectrum of the SWCNTs film before and after
the subtraction of the linear background (dashed line).} \label{fgr:freestanding}
\end{figure}

The optical absorbance spectrum of the free-standing film of SWCNTs is depicted
in the inset of Fig.\ \ref{fgr:freestanding} (a). It consists of several pronounced absorption bands on top of a broad background, which
is due to the $\pi$-$\pi$$^*$ absorption centered at around 5~eV. We approximate the background by a linear function [see inset of Fig.~\ref{fgr:freestanding} (a)].
After subtracting the background the pronounced absorption peaks are more clearly seen,
as illustrated in the inset of Fig.\ \ref{fgr:freestanding} (a), showing the optical absorbance spectrum before and after the background subtraction as well as the linear background. The bands labeled S$_{ii}$ and M$_{ii}$ correspond to the $i$th optical transitions in semiconducting and metallic SWCNTs, respectively.
The optical transitions S$_{11}$, S$_{22}$, and M$_{11}$ exhibit a fine structure, which reflects the nanotube diameter distribution in the sample. Therefore, several Lorentz oscillators are needed to describe each band. By fitting
the absorbance spectrum with Lorentz functions the frequencies of the various transitions were extracted and are given in Table \ref{table:IR}. We plot in Fig.\ \ref{fgr:freestanding} (a) the total fitting curve and the various Lorentz contributions in addition to the optical absorbance data, in order to illustrate the analysis procedure.

\begin{table}[b]
\caption{Peak position frequencies (in cm$^{-1}$) of the Lorentzian contributions from
the different optical absorption bands for SWCNTs and DWCNTs (error bar $\pm$10~cm$^{-1}$ for S$_{11}$ and S$_{22}$; error bar $\pm$30~cm$^{-1}$ for M$_{11}$, S$_{33}$, and S$_{44}$).}
\label{table:IR}
\begin{tabular}{cccc}\hline \hline
\multirow{1}{*}{}
$\vspace{5pt}$
  S$_{11}$ & S$_{22}$ & M$_{11}$ & S$_{33/44}$\\ \hline
\multirow{4}{*}{}
 $\vspace{2pt}$
 \textit{SWCNTs}& & & \\
  4800 & $\hspace{20pt}$ 8960 & $\hspace{20pt}$ 13300 & $\hspace{20pt}$ 17763 \\
  5255 & $\hspace{20pt}$ 9610 & $\hspace{20pt}$ 14180 & $\hspace{20pt}$ 18873 \\
  6920 & $\hspace{20pt}$ 10330& $\hspace{20pt}$ 15098 & $\hspace{20pt}$ 19909 \\ \hline
\multirow{4}{*}{}
 $\vspace{2pt}$
 \textit{DWCNTs}& & & \\
 ....... & $\hspace{20pt}$ 8930 & $\hspace{20pt}$    13270  & $\hspace{20pt}$ 17700 \\
  5185   & $\hspace{20pt}$ 9580 & $\hspace{20pt}$    14100  & $\hspace{20pt}$ 18780 \\
  6870   & $\hspace{20pt}$ 10300& $\hspace{20pt}$ ......... &$\hspace{20pt}$ 19835 \\ \hline
 \end{tabular}
\end{table}

The above-described analysis was also applied to the optical absorbance spectrum of the free-standing
DWCNT film. The resulting curve is plotted in Fig.\ \ref{fgr:freestanding} (b) after the background subtraction together with the fitting curve and the various Lorentz terms. The so-extracted frequencies of the Lorentz oscillators are included in Table \ref{table:IR}.
Overall, the features in the spectrum resemble those of SWCNTs, but the number of contributions to the S$_{11}$ and M$_{11}$ absorption bands is smaller. Importantly, in the DWCNT absorbance spectrum no new features appear, which could be attributed to optical transitions in the inner tubes, in agreement with an earlier report.\cite{Botka10} This can be explained by the overlap of the optical absorption features related to the inner tubes with those related to the outer tubes. According to the Kataura plot \cite{Kataura99} the energies of the S$_{11}$ transitions in the inner tubes with $d$$\sim$0.8~nm are close to those of the S$_{22}$ transitions in the outer tubes with $d$$\sim$1.4~nm, while the S$_{22}$ transitions of the inner tubes approximately coincide with the S$_{33}$ transitions of the outer tubes. However, a comparison of the DWCNT absorbance spectrum with that of SWCNTs reveals no new bands, which could be unambiguously attributed to optical transitions in the inner tubes of DWCNTs.

Comparing the parameters of corresponding Lorentz contributions in the SWCNT and the DWCNT film, one notices a small but significant redshift of all contributions in case of the DWCNT film. This redshift can be attributed to the slightly enlarged average diameter of the outer tubes in the DWCNTs as compared to the SWCNTs' average tube diameter,\cite{Kataura99} revealed by Raman results.\cite{Anis12}

\subsection{Optical spectroscopy of SWCNTs and DWCNTs under pressure}

\begin{figure}
\includegraphics[width=0.9\columnwidth]{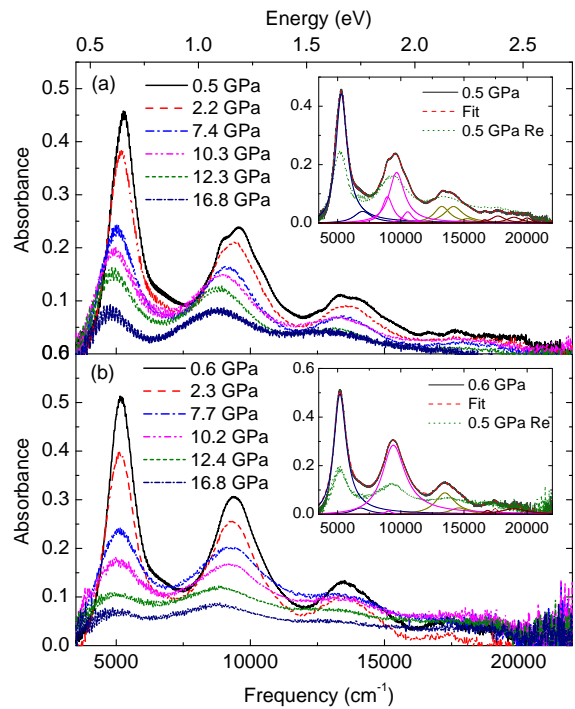}
\caption{Background-subtracted optical absorbance spectra of the (a) SWCNTs and (b) DWCNTs films as a function of pressure. The insets in (a) and (b) depict the optical absorbance spectrum at the lowest pressure during pressure increase together with the various Lorentzian contributions (using the same color code as for the fits of Fig.\ \ref{fgr:freestanding}), and the absorbance spectrum at the lowest pressure during pressure release, for the SWCNT and DWCNT films respectively.}
\label{fgr:Nitrogen1}
\end{figure}

The mechanical stability of the bundled SWCNTs and DWCNTs were probed by transmittance measurements on the prepared films under hydrostatic pressure. The corresponding pressure-dependent optical absorbance spectra are depicted in Fig.~\ref{fgr:Nitrogen1}, after the background subtraction. Already at the lowest applied pressure ($\sim$0.5~GPa) the absorption bands are broadened and the fine-structure due to different tube diameters is smeared out, as compared to the free-standing films. Therefore, a smaller number of Lorentz contributions is needed for obtaining a good fit of the spectra. This can be seen in the insets of Fig.~\ref{fgr:Nitrogen1}(a) and (b) showing the optical absorbance spectra of SWCNT and DWCNT films, respectively, at the lowest pressure together with the total fitting curve and the Lorentz terms. In the further analysis and discussion only the strong Lorentz contributions will be considered. In case of SWCNTs we observe only one strong contribution for the S$_{11}$ band and two strong contributions for S$_{22}$ and M$_{11}$ bands, which are marked by S$_{22}$(1), S$_{22}$(2) and by M$_{11}$(1), M$_{11}$(2), respectively. For DWCNTs we observe only one strong contribution for the S$_{11}$, S$_{22}$, and M$_{11}$ absorption bands.

One notices a redshift of all absorption bands for both SWCNT and DWCNT films under pressure application. This redshift is consistent with earlier observations \cite{Thirunavukkuarasu10,Kuntscher10} and is ascribed to $\sigma^*$-$\pi^*$ hybridization and symmetry breaking.\cite{Wu04,Charlier96,Liu08} Besides the redshift, the absorption bands broaden and lose spectral weight with increasing pressure: In the case of the SWCNT film all optical transitions S$_{11}$, S$_{22}$, and M$_{11}$ are resolvable up to the highest applied pressure (16.8~GPa), but for the DWCNT film especially the M$_{11}$ band is smeared out considerably at 16.8~GPa.

\begin{figure}
\includegraphics[width=0.9\columnwidth]{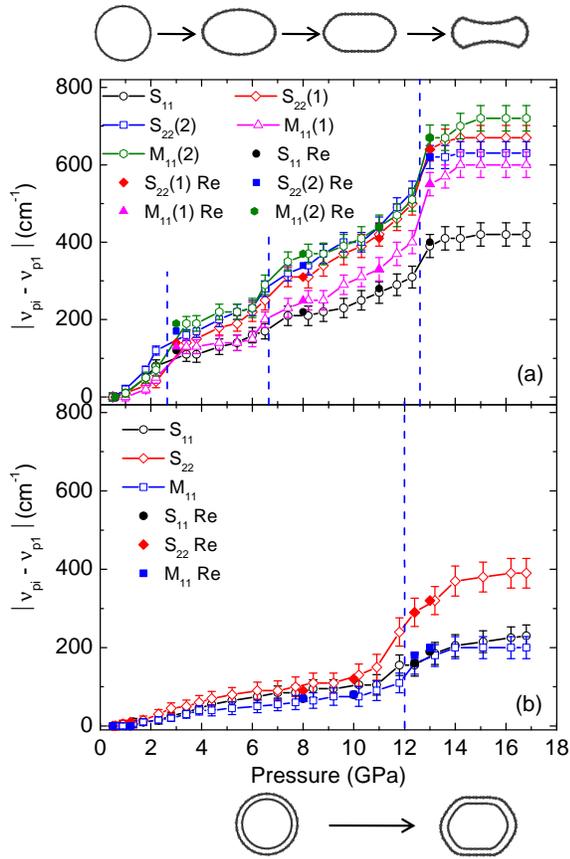}
\caption{Relative energy shift of the optical transitions as a function of pressure for (a) SWCNTs and (b) DWCNTs films. The shift was calculated as the difference between the absorption frequency of the contribution $\nu$$_{pi}$ at a pressure $pi$, and the absorption frequency $\nu$$_{p1}$ at the lowest pressure value $p1$. All the contributions during pressure release are marked as  S$_{ii}$ (Re) and M$_{ii}$ (Re). The vertical, dashed lines mark the critical pressure values. The proposed deformation of the cross-sections for SWCNTs and DWCNTs in the different pressure ranges is illustrated.}
\label{fgr:Nitrogen2}
\end{figure}

For a quantitative analysis of the pressure-induced redshift of the absorption bands, we plot in Fig.\ \ref{fgr:Nitrogen2} the relative energy shift of the Lorentz contributions as a function of applied pressure.
For the SWCNT film [see Fig.~\ref{fgr:Nitrogen2}(a)] the pressure-induced shifts show a steep increase at the critical pressures P$_{c1}$$\approx$3~GPa, P$_{c2}$$\approx$7~GPa, and P$_{c3}$$\approx$13~GPa, followed by a plateau. According to earlier results and theoretical predictions,\cite{Thirunavukkuarasu10,Capaz04,Chan03,Sluiter04,Elliott04,Hasegawa06,Venkateswaran99,Peters00,Sandler03,Gadagkar06} we interpret the anomaly at P$_{c1}$ in terms of a structural phase transition, where the tube is deformed from a circular to an oval shape. Above P$_{c2}$ a more drastic change in the cross section from oval to race-track or peanut-type shape occurs.\cite{Yao08,Caillier08} The plateau with an onset at P$_{c3}$ indicates a saturation of the pressure-induced deformation above this critical pressure. According to recent pressure investigations on SWCNTs with an average tube diameter of 1.35~nm, the transition pressure for the collapse of the tubes amounts to $\sim$14~GPa.\cite{Aguiar11,Caillier08} Since the average tube diameter in our SWCNT sample is slightly larger, a slightly smaller transition pressure is to be expected. Hence, we interpret the onset of the plateau at P$_{c3}$ in terms of the deformation to a peanut shape (collapse). The proposed sequence of structural deformations in SWCNTs is illustrated on top of Fig. \ref{fgr:Nitrogen2}.
Interestingly, stabilization effects due to nitrogen filling of the SWCNTs are not found in our data. Furthermore, we obtain qualitatively similar results (not shown) when using argon as pressure transmitting medium. In conclusion, neither nitrogen nor argon filling of the SWCNTs prevents pressure-induced anomalies in the electronic transitions due to tubular deformation, in contradiction to earlier reports.\cite{Merlen05}

The relative energy shifts of the optical transitions S$_{11}$, S$_{22}$, and M$_{11}$ in the DWCNT bundles are depicted in
Fig.~\ref{fgr:Nitrogen2}(b). The first important result is that the redshift of absorption bands in DWCNTs is small compared to the SWCNTs, and within the error bars an approximately linear behavior with a pressure coefficient in the range $\sim$9-15~cm$^{-1}$/GPa is found. For SWCNTs in comparison, the relative (non-linear) energy shift of the bands as a function of pressure is significantly larger and has
reached large values already at P$_{c1}$. The smaller redshift of the absorption bands in DWCNTs indicates that pressure-induced $\sigma^*$-$\pi^*$ hybridization and symmetry breaking effects are reduced in the outer tubes of the DWCNTs. Hence, already at low pressures, i.e., before a structural transition takes place, the outer tube is mechanically stabilized by the inner tube. This finding is in contrast to earlier results obtained by Raman studies which probe the vibrational properties: Here, the pressure coefficient of the tangential optical phonon mode for the outer tube was found to be similar to the one in SWCNTs, and also severe pressure-induced fading and broadening of the tangential and radial breathing modes of the outer tubes were observed.\cite{Puech04,Arvanitidis05,Arvanitidis05a} Thus, in these earlier investigations stabilization effects by the inner tubes were not obvious in the low-pressure regime.

Above $\sim$12~GPa the relative energy shifts of the absorption bands in the DWCNT bundles steeply increase, followed by a plateau above $\sim$14~GPa [see Fig.~\ref{fgr:Nitrogen2}(b)]. This anomalous behavior signals the onset of strong pressure-induced alterations in the electronic properties of the outer tube at a critical pressure P$_d$$\approx$12~GPa related to tubular deformation. The pressure P$_d$ is a factor of $\sim$4 higher than the first critical pressure P$_{c1}$ in SWCNTs with a similar average diameter.
Several earlier experimental studies (mostly Raman) reported higher critical pressures for the structural transition of the outer tube in DWCNTs as compared to SWCNTs, and this was attributed to the structural support of the outer tube by the inner tube.\cite{Gadagkar06,You11,Aguiar11} The enhanced structural stability of the outer tube was theoretically predicted by Yang et al. \cite{Yang06}:
For small-diameter (5,5)\@(10,10) DWCNT bundles ($d_{inner}$$\approx$0.68~nm, $d_{outer}$$\approx$1.36~nm) a small discontinuous volume change appears at the critical pressure P$_d$=18~GPa, accompanied by a cross sections' change between two deformed hexagons; the collapse to peanut shaped cross sections, however, happens at higher pressure.
In contrast, (7,7)\@(12,12) DWCNT bundles ($d_{inner}$$\approx$0.95~nm, $d_{outer}$$\approx$1.63~nm) undergo one structural phase transition and collapse at P$_d$=10.6~GPa. Because of the similar diameters, our DWCNTs are expected to show similar effects under pressure like the (5,5)\@(10,10) DWCNTs, but with a somewhat smaller value of the critical pressure.

According to molecular dynamics simulations by Gadagkar et al.\cite{Gadagkar06} the tubes cross sections remain nearly circular up to a critical pressure, where the tube cross sections are deformed to an elliptical shape. The critical pressure follows a 1/$R_{eff}^3$ dependence,\cite{Gadagkar06} where the effective radius of the DWCNT
$R_{eff}$ is defined as $1/R_{eff}^3=(1/n)\sum_{i=1}^n(1/R_{i}^3)$ with n=2 for DWCNTs.
Based on this relation we calculate
the effective radius for our DWCNT bundles with d$_{inner}$$\approx$0.80~nm and  d$_{outer}$$\approx$1.45~nm, as estimated from the HRTEM images and Raman spectroscopy,\cite{Anis12}
to $R_{eff}$$\approx$0.5~nm and obtain a critical pressure value of P$_d$$\approx$11~GPa according to Ref.~\onlinecite{Gadagkar06}.
Thus, our experimental finding of a critical pressure P$_d$$\approx$12~GPa is in very good agreement with theoretical predictions for the onset of the deformations of the tube cross-sections in DWCNTs bundles. Corresponding experiments on DWCNTs using argon as pressure transmitting medium (not shown) give qualitatively similar results.

Finally, we comment on the reversibility of the pressure-induced structural changes.
For both SWCNTs and DWCNTs the pressure-induced frequency shifts of the optical transitions
are reversible upon pressure release [see Figs.~\ref{fgr:Nitrogen2}(a) and (b)].
However, one notices a loss in intensity, namely up to $\sim$50\% and $\sim$60\% of the original absorbance value for SWCNTs and DWCNTs, respectively. This is illustrated in the insets of Fig.~\ref{fgr:Nitrogen1}(a) and (b), where the optical absorbance spectra at the lowest pressure during pressure release are included.
The irreversible changes indicate that a fraction of the tubes has been permanently damaged during pressure loading, with a loss of the characteristic features in their density of states.

\section{Conclusions}

In conclusion, the mechanical stability of the outer tubes in DWCNTs under hydrostatic pressure has been characterized in terms of the optical transitions. Compared to empty nanotubes, the redshift of the absorption bands in DWCNTs is smaller and follows an approximately linear behavior with a pressure coefficient in the range $\sim$9-15~cm$^{-1}$/GPa. The reduced redshift indicates that the outer tube is stabilized by the inner tube regarding its electronic properties. Anomalies in the pressure-induced shifts of the absorption bands in DWCNTs at around P$_d$$\approx$12~GPa signal the onset of a deformation of the outer tubes.
According to theoretical predictions we interpret the anomalies in terms of a small discontinuous volume change accompanied by a cross sections' change to two deformed hexagons. The pressure-induced alterations of the absorption bands are reversible regarding their frequency position but not completely reversible regarding their intensity, indicating that a fraction of the tubes has been permanently damaged under high pressure.

\section{Acknowledgment}
We acknowledge financial support by the German Science Foundation (DFG), the DAAD, and the Egyptian Government.
{}

\end{document}